\documentclass[preprint,12pt,Arial,showpacs,showkeys]{revtex4}

\usepackage{color}
\usepackage{amssymb}
\usepackage{amsmath}
\usepackage{graphicx}
\usepackage{graphics}
\usepackage{dcolumn}
\usepackage{bm}
\usepackage{color}
\usepackage{epstopdf}
\usepackage{hyperref}

\providecommand{\keywords}[1]{\textbf{\textit{Index terms---}} #1}

\begin{document}
	
	\newcommand{\blue}[1]{\textcolor{blue}{#1}}
	\newcommand{\ans}{\blue}
	
		
		
		
		\title{Logarithmic Regge Pole}
		
		
		\author{S. D. Campos\email{sergiodc@ufscar.br}}
		
		\affiliation{Applied Mathematics Laboratory-CCTS/DFQM, Federal University of S\^ao Carlos, Sorocaba, S\~ao Paulo CEP 18052780, Brazil}
		
		\begin{abstract}
		This work presents the subtraction procedure and the Regge cut in the logarithmic Regge pole approach. The subtraction mechanism leads to the same asymptotic behavior as previously obtained in the non-subtraction case. The Regge cut, on the other hand, introduces a clear role to the non-leading contributions for the asymptotic behavior of the total cross section. From these results, one introduces some simple parameterization to fit the experimental data for the proton-proton and antiproton-proton total cross section above some minimum value up to the cosmic-ray. The fit parameters obtained are used to present predictions for the $\rho(s)$-parameter as well as to the elastic slope $B(s)$ at high energies.  
		\end{abstract}
		

			
			
		
	
	





\maketitle

\section{Introduction}\label{sec:intro}

The introduction of the complex angular momentum $l$ in the potential scattering gives rise to poles representing bound states or resonances for $l>0$ \cite{t.regge.nuovo.cim.14.951.1959,t.regge.nuovo.cim.18.947.1960,A.Bottino.A.M.Longoni.T.Regge.Nuovo.Cim.23.954.1962}. As well-known, the application of Regge's original ideas to the high energy scattering of elementary particles results in a very successful theory. For example, the broad class of reactions that can be explained in a unified view led Donnachie and Landshoff to conjecture that the Regge theory would be part of the truth of the particle physics  \cite{A.Donnachie.P.V.Landshoff.Phys.Lett.B.296.227.1992}. In other words: if such formalism is wrong, then it is wrong, probably, by some little misunderstanding. 

The main entity in the Regge theory is the $\alpha(t)$ trajectory, which is a phenomenological input within the formalism. The location of a pole corresponds to a bound state, and the scattering amplitude behavior is controlled by this leading Regge pole. Then, a resonance with spin $j$ turns necessary to include the excitation $j+2$, $j+4$,..., keeping fixed the other quantum numbers. These particles lies on the Regge trajectory, obtained from the so-called Chew-Frautschi plot \cite{G.F.Chew.S.C.Frautschi.Phys.Rev.Lett.7.394.1961,G.F.Chew.S.C.Frautschi.Phys.Rev.Lett.8.41.1962}. The Regge trajectory is assumed to be linear for light mesons and baryons \cite{A.J.G.Hey.R.L.Kelly.Phys.Rep.96.71.1983}, whose parameters can be extracted from the scattering data as well as from the particle spectra. However, an important question without an answer is if these trajectories are linear everywhere or only at the asymptotic regime. Moreover, the analyticity and unitarity properties require that $\alpha(t)$ be a non-linear complex-valued function \cite{A.Degasperis.E.Predazzi.Nuovo.Cimento.A.65.764.1970,A.I.Bugrij.G.Cohen.Tannoudji.L.L.Jenkovszky.N.A.Kobylinsky.Fortschritte.der.Physik.21.427.1973}. 

The Pomeranchuk theorem asserts the total cross section of particle-particle and particle-antiparticle at high energies should tend to the same limit \cite{I.Ia.Pomeranchuk.Sov.Phys.JETP.7.499.1958}. This theorem creates the need for the exchange of a state, with the quantum numbers of the vacuum, that not differs particles from antiparticles in the asymptotic energy regime. This is the role played by the leading Regge pole, the so-called pomeron, in the particle scattering at high energies, and the reason supporting its existence is entirely phenomenological. The problem here is that $\alpha(t)$ for the pomeron can only be obtained from the experimental data since it has no particle spectrum. However, if one extrapolates the pomeron trajectory from negative to positive values, then one can predict glueball states with integer spin. The first mention to the pomeron as a pair of gluons in a color singlet is due to Low and Nussinov \cite{F.E.Low.Phys.Rev.D.12.163.1975,S.Nussinov.Phys.Rev.Lett.34.1286.1975}. Nonetheless, there is no experimental evidence for the pomeron in present-day energies. 

The odderon, the counterpart of the pomeron, can distinguish particles from antiparticles in the convenient energy range. The original idea of the odderon is due to Bouquet \textit{et al.} \cite{A.Bouquet.B.Diu.E.Leader.B.Nicolescu.Nuovo.Cimento.29A.30.1975} and Joynson \textit{et al.} \cite{D.Joynson.E.Leader.C.Lopez.B.Nicolescu.Nuovo.Cimento.30A.345.1975} and the proposition of the odderon as the exchange of three reggeized gluons is due to Bartels \cite{J.Bartels.Nucl.Phys.B175.365.1980}, followed close by Refs. \cite{J.Kwiecinski.M.Praszalowicz.Phys.Lett.B94.413.1980, T Jaroszewicz.J.Kwiecinski.Z.Phys.C12.167.1982}, being a formal QCD prediction. On the other hand, considering the $t$-dependence of the differential cross section, the general belief is that the odderon is responsible by the pronounced dip in the proton-proton scattering. On the other hand, it fills the dip in the case of the proton-antiproton differential cross section. Differently from the pomeron, the odderon has a possible experimental evidence \cite{G.Antchev.etal.TOTEM.Coll.Eur.Phys.J.C79.785.2019} subject to discussions in the literature \cite{E.Martynov.B.Nicolescu.Phys.Lett.B778.414.2018,V.A.Khoze.A.D.Martin.M.G.Ryskin.Phys.Lett.B780.352.2018,A.Szczurek.P.Lebiedowicz.PoS.DIS2019.071.2019}.

Since a long time ago, one knows that the Regge theory is valid in the perturbation theory. Indeed, the Regge trajectory also can be obtained from the Bethe-Salpeter equation \cite{B.W.Lee.R.F.Sawyer.Phys.Rev.127.2266.1962}. The BFKL equation also result in the pomeron \cite{V.S.Fadin.E.A.Kuraev.L.N.Lipatov.Sov.Phys.JETP44.443.1976,Y.Y.Balitsky.L.N.Lipatov.Sov.J.Nucl.Phys.28.822.1978}. This leading Regge pole emerging in the perturbative QCD approach is called hard pomeron, and it is used to describe the behavior at a small $x$ (the Bjorken scale) in deep inelastic scattering as well as in diffractive processes. The non-perturbative leading Regge pole, in contra-position, is called soft pomeron. Efforts towards a unified view of both pomeron pictures are being conduced \cite{J.Bartels.C.Contreras.G.P.Vacca.JHEP.01.004.2019}.

However, not everything is a flower in Regge's garden. The Froissart-Martin (FM) bound, for example, disagree with the rise of the total cross section given by the leading Regge pole \cite{m.froissart.phys.rev.123.1053.1961,a.martin.nuovo.cim.42.930.1966}. The Regge theory predicts the total cross section asymptotically behaving as $s^{\alpha(0)-1}$, as $s\rightarrow\infty$. The FM bound predicts, however, a rising bounded by $\ln^2 s$. The only way to ensure the validity of the Regge formalism in front of the FM bound is with a trajectory of less than 1. Nonetheless, the fitting procedures for the total cross section always produces $\alpha(0)>1$ \cite{A.Donnachie.P.V.Landshoff.Phys.Lett.B.296.227.1992,R.C.Badatya.P.K.Patnaik.Pramana.15.463-474.1980}. The FM bound is a crucial formal result of the high energy physics and cannot be disregarded in any theoretical approach. The analyticity principle is also not be satisfied unless the trajectory of all particles lies on the Regge trajectory.

Recently, obeying the FM bound, a novel approach to the leading Regge pole was obtained by introducing a logarithmic representation for the leading Regge pole \cite{S.D.Campos.Phys.Scrip.xx.2020}. In the present work, one continues the logarithmic Regge approach introducing the subtraction and the cut problem in the logarithmic Regge framework.

The subtraction procedure, in the present formalism, cannot be used in its \textit{pure version} since the fast decreasing caused by the subtraction $s^{-1}$ turns the approach useless. However, as shall be seen, there is a subtle approximation allowing the use of a less restrictive version of the subtraction mechanism. This approach will lead to a modified subtraction mechanism, whose consequence is that subtraction and non-subtraction cases produce the same functional form to the asymptotic scattering amplitude. The cut, on the other hand, seems to be a result of the sub-leading contributions. Then, this mechanism may be particularly important to explain the mixed energy region where the total cross section, for example, is controlled by the pomeron and odderon exchange.

Using naive parameterizations for the proton-proton and proton-antiproton total cross section, it is possible to understand the role of the pomeron at high energies within the logarithmic Regge approach. As obtained in \cite{S.D.Campos.Phys.Scrip.xx.2020}, the double pomeron picture is favored for energies above 1.0 TeV. On the other hand, as shall be seen, starting the fitting procedures taking into account energies above $25.0$ GeV, then the pomeron assumes values greater than 2 suggesting the saturation of the Froissart-Martin bound. This problem can be solved by using the recent TOTEM measurement of $\rho(s)$ \cite{G.Antchev.etal.TOTEM.Coll.Eur.Phys.J.C79.785.2019}.

Applying the derivative dispersion relation, one obtains the real part of the elastic scattering amplitude. The parameters are obtained from the fitting procedures, and predictions for the $\rho(s)$-parameter are presented. Assuming a null subtraction constant (it interferes only in the low energy experimental data), the curves suggest a double-pomeron exchange to reproduce the $\rho(s)$ obtained in the TOTEM Collaboration. Then, using this constraint, one keeps the double-pomeron trajectory, obtaining a general description of the total cross section obeying the FM bound, and the correct prediction for the $\rho(s)$-parameter at high energies. Hereafter, $\rho(s)=\rho$.

The paper is organized as follows. Section \ref{sec:fr} presents the experimental data set and the fitting procedure. In the section \ref{sec:rgt}, one presents the main results of \cite{S.D.Campos.Phys.Scrip.xx.2020} and develops the subtracted case as well as the Regge cut case. Section \ref{sec:dr} presents a brief discussion about the $\rho$-parameter as well as predictions for the slope of the differential cross section. The critical remarks are the subject of the last section \ref{sec:critical}.

\section{Experimental Data and Fitting Procedures}\label{sec:fr}

The main quantity in the forward elastic scattering is the total cross section, connected with the imaginary part of the forward elastic scattering amplitude through the optical theorem. Bearing this in mind, one considers here the experimental data for the proton-proton ($pp$) and proton-antiproton ($p\bar{p}$) total cross sections, $\sigma_{tot}^{pp}(s)$ and $\sigma_{tot}^{p\bar{p}}(s)$. As usual, $t$ is the squared momentum transfer and $s$ is the squared energy, both in the center-of-mass system.

\begin{figure}
	\centering{
		\includegraphics[scale=0.45]{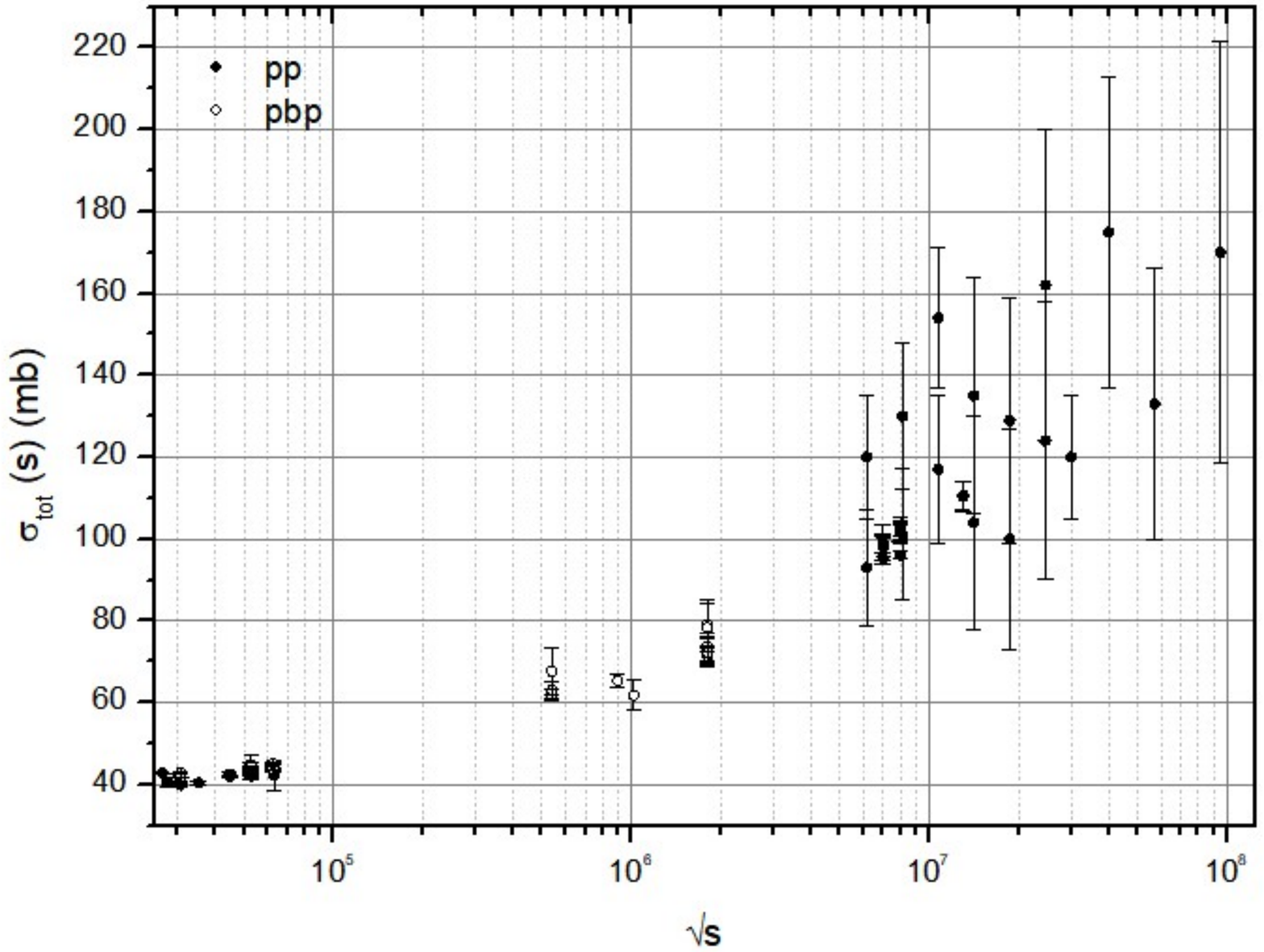}
		\vspace{-0.5cm}
		\caption{Experimental data for the $pp$ and $p\bar{p}$ total cross sections above $\sqrt{s}\geq 25.0$ GeV up to the cosmic-ray. Experimental data are from \cite{PDG-PhysRev-D98-030001-2018,G_Antchev_TOTEM_ Coll_Eur_Phys_J_C79_103_2019}.}
		\label{fig:fig_0}}
\end{figure}

These $pp$ and $p\bar{p}$ experimental data are used to form a joint data set since the Pomeranchuk theorem asserts they tend to the same limit if $s\rightarrow\infty$. This behavior, predicted to occur only at the asymptotic regime, seems yet to be started from energies above $\sqrt{s}\geq 25.0$ GeV. Figure \ref{fig:fig_0} shows the experimental data used in the fitting procedures for $\sigma_{tot}^{pp}(s)$ and for $\sigma_{tot}^{p\bar{p}}(s)$.

The goal of treating both data set as only one resides on the possibility that the absence of $pp$ experimental data, on some energy range, can be compensated by the existence of $p\bar{p}$ data in this range, and vice-versa. Moreover, there is no data selection in any set considered. The following experimental data set are used throughout this paper.

\begin{itemize}
	
\item The SET 1 is formed by the experimental data for $\sigma_{tot}^{pp}(s)$ and $\sigma_{tot}^{p\bar{p}}(s)$  above $\sqrt{s}=1.0$ TeV up to the cosmic-ray data. 

\item The SET 2 uses the experimental data for $\sigma_{tot}^{pp}(s)$ and $\sigma_{tot}^{p\bar{p}}(s)$ above $\sqrt{s}=1.0$ TeV, excluding the cosmic-ray experimental data. 

\item The SET 3 contains experimental data for $\sigma_{tot}^{pp}(s)$ and $\sigma_{tot}^{p\bar{p}}(s)$  above $\sqrt{s_c}$ GeV up to the cosmic-ray data.

\item The SET 4 excludes the cosmic-ray data from the SET 3.

\end{itemize}

The energy cut $\sqrt{s_c}$, corresponds to the energy for which the total cross section achieves its minimum value \cite{S.D.Campos.C.V.Moraes.V.A.Okorokov.Phys.Scrip.2019}. For the SET 1 and 2, one uses $\sqrt{s_c}=25.0$ GeV. However, as shall be seen, the emergence of a term $\ln\ln(s/s_c)$ in the logarithmic Regge cut implies a change in $\sqrt{s_c}$ to avoid the emergence of negative or complex values. Then, for the SET 3 and 4, one uses $\sqrt{s_c}=15.0$ GeV, which is a consequence of the constraint $\sqrt{s}\geq\sqrt{e s_c}$, where $e$ is the Napier's constant.  

Using the derivative dispersion relations, one obtains the real part of the forward elastic scattering amplitude. Then, it is possible to present the predictions for the $\rho$-parameter based on the fitting results. One uses only the experimental data for $\rho$ ($pp$ and $\bar{p}p$) above $\sqrt{s}=25.0$ GeV.  It is important to stress these experimental values were not used in the fitting procedures, i.e. they are used only as a background to show the predictions for the $\rho$-parameter. It is also shown predictions for the slope of the differential cross section considering some energies of interest. Moreover, a fitting procedure for the slope using the $pp$ and $\bar{p}p$ experimental data above 1.0 TeV is performed.

All the experimental data were collected from Particle Data Group \cite{PDG-PhysRev-D98-030001-2018}. Moreover, $\sigma_{tot}^{pp}(s)$ at $\sqrt{s}=2.76$ TeV is from \cite{G_Antchev_TOTEM_ Coll_Eur_Phys_J_C79_103_2019}. Hereafter, one uses only $\sigma_{tot}(s)$ to refer to both $\sigma_{tot}^{pp}(s)$ and $\sigma_{tot}^{p\bar{p}}(s)$ (the same for $\rho$).  


\section{The logarithmic leading Regge Pole}\label{sec:rgt}

In the Regge theory, the scattering amplitude is written as an analytic function of the angular momentum $J$. This representation is formulated in the high energy limit $s\rightarrow\infty$, and associates the asymptotic behavior of the scattering amplitude in the $s$-channel to the exchange of one-particle or more, represented in the $t$-channel by the leading Regge poles. One writes the scattering amplitude as
\begin{eqnarray}
A(s,t)=\mathrm{Re}A(s,t)+i\mathrm{Im}A(s,t),
\end{eqnarray}

\noindent and one also assumes that behavior of such a function, at very high energies, is given only by the absorptive part $A(s,t)\approx\mathrm{Im}A(s,t)$. As well-known, in the usual Regge pole formalism, one writes the asymptotic scattering amplitude as
\begin{eqnarray}\label{eq:regge_pole}
A(s,t)\rightarrow (\eta+e^{-i\pi\alpha(t)})\beta(t)(s/s_c)^{\alpha(t)}, ~~~s\rightarrow\infty,
\end{eqnarray}

\noindent where $\eta=\pm 1$ is the signature related with the crossing symmetry $s\leftrightarrow t$, $\sqrt{s_c}$ some critical energy, and $\beta(t)$ is the residue function of the pole depending only on $t$. In this formalism, $s$ and $t$ are treated as complex-valued functions. Using (\ref{eq:regge_pole}), one can attain the asymptotic form of the differential cross section 
\begin{eqnarray}\label{eq:dif_1}
\frac{d\sigma}{dt}\approx (s/s_c)^{2\alpha(t)},
\end{eqnarray}

\noindent and adopting the normalization $s\sigma_{tot}(s)=\mathrm{Im}A(s,t)\approx A(s,t)$, one has
\begin{eqnarray}\label{eq:totalregge}
\sigma_{tot}(s)\approx (s/s_c)^{\alpha(0)-1}.
\end{eqnarray}

The mathematical disagreement between (\ref{eq:totalregge}) and the FM bound given below
\begin{eqnarray}\label{eq:froissart}
\sigma_{tot}(s)\leq c\ln^2(s/s_c),
\end{eqnarray}

\noindent where $c$ is some real constant, is inevitable for $1<\alpha(0)$. However, it is important to stress that the Regge theory predicts both a leading pole behaving as $(s/s_c)^{\alpha(t)}$ as well as sub-leading terms behaving as $\ln^k (s/s_c)$. If $\alpha(0)=1$ in (\ref{eq:totalregge}), only the logarithmic terms survive, being controlled by the FM bound, which means $k\leq 2$. Therefore, the precise knowledge of the leading pole functional form in the Regge approach is subject to the experimental evidence.

However, the Regge theory and the FM bound can agree with each other if it is imposed a constraint on the scattering angle as well as a mathematical approximation on the cosine series \cite{S.D.Campos.Phys.Scrip.xx.2020}. The scattering angle is restricted to the range $0\leq \cos \theta \leq 1$ for $|t|<\!\!< s$. In this case, one writes in the $s$-channel the following approximation, valid near the forward direction, for the cosine of the scattering angle \cite{S.D.Campos.Phys.Scrip.xx.2020}
\begin{eqnarray}\label{eq:cos_2}
\cos (\theta) = 1+\frac{2t}{s} \lesssim \ln \left(1+\sqrt{e}\left(1+\frac{2t}{s}\right)\right),
\end{eqnarray}

\noindent where physical $t$ is negative, $s>4m^2$, and $m$ is particle mass. It is important to remind that, in the $s$-channel,  fixed $t$ represents the squared momentum transfer, and $s$ is the squared energy. Moreover, $|\cos\theta|\leq 1$, turning necessary to use the crossing property $s\leftrightarrow t$ in the high energy limit $s\rightarrow\infty$.

The approximate representation (\ref{eq:cos_2}) can be obtained by noting that usual cosine series can be written
\begin{eqnarray}\label{cos_1}
	\cos(x)=1-\sum_{n=1}^{\infty} \frac{(-1)^{n-1}}{(2n)!}x^{2n},
	\end{eqnarray}

\noindent where the sum in the r.h.s enclose the information about the energy and momentum transfer. Observe the following inequalities holds for $n\in \mathbb{N}$
\begin{eqnarray}\label{inel}
	(2n)^{2n+\frac{1}{2}}\geq n^{n+\frac{1}{2}}\geq n^n\geq n.
	\end{eqnarray}

Then, the factorial number in (\ref{cos_1}) can be circumvented by using the Stirling's approximation
\begin{eqnarray}\label{stirling}
	(2n)!\approx (2n)^{2n+\frac{1}{2}}e^{-2n}\sqrt{2\pi}.
	\end{eqnarray}

Then,
\begin{eqnarray}\label{cos_2}
	\cos(x)=1-\sum_{n=1}^{\infty} \frac{(-1)^{n-1}}{(2n)!}x^{2n}\approx 1-\sum_{n=1}^{\infty} \frac{(-1)^{n-1}e^{2n}}{(2n)^{2n+\frac{1}{2}}\sqrt{2\pi}}x^{2n},
	\end{eqnarray}

To reduce the series on the r.h.s. of (\ref{cos_1}) into the logarithmic series, one notices that using (\ref{inel}) one can always find a real number $a$, satisfying
\begin{eqnarray}\label{inel_2}
\frac{e^{2n}}{\sqrt{2\pi}(2n)^{2n+\frac{1}{2}}}\leq \frac{a^{2n}}{n}.
\end{eqnarray}
	
Of course, the result (\ref{inel_2}), when replaced in the convenient series, possibly implies a slower convergence than the original cosine series, near the forward direction. Moreover, the choice of $a$ is not unique. For example, the above inequality holds for $a=e=2.71828...$, where $e$ is the Napier's number.

Using the results (\ref{stirling}) and (\ref{inel_2}), one can exchange the series on r.h.s. of (\ref{cos_1}) by the approximation
\begin{eqnarray}\label{cos_2}
\cos(x)\lesssim 1- \sum_{n=1}^{\infty}\frac{(-1)^{n-1}}{n}[(ax)^2]^n=1-\sum_{n=1}^{\infty}\frac{(-1)^{n-1}}{n}[(y)]^n,
	\end{eqnarray}
	
\noindent with the condition $y\geq 0$. Using the last result, one finally obtains
\begin{eqnarray}\label{final_1}
	1-\sum_{n=1}^{\infty}\frac{(-1)^{n-1}}{n}[(y)]^n=\ln\left(\frac{e}{1+y}\right).
	\end{eqnarray}
	
To ensure the first-order approximation, one uses
\begin{eqnarray}\label{usoy}
	y=e-\left[1+\sqrt{e}\left(1+\frac{2t}{s}\right)\right],
	\end{eqnarray}
	
\noindent where the factor $\sqrt{e}$ comes from the fact that $y\geq0$. Observe that, for $t=0$, one has $\cos\theta= 1$, and the approximation performed furnish $\cos\theta\approx 0.97$.

Using the asymptotic properties of the Legendre polynomial, taking into account the approximation (\ref{eq:cos_2}), and the crossing $s\leftrightarrow t$, one writes $(s\rightarrow\infty$ and fixed $t$)
\begin{eqnarray}\label{eq:asymp_1}
P_l(s)\rightarrow \ln^{l} (s/s_c).
\end{eqnarray}

Indeed, the above result can be used to write the asymptotic scattering amplitude as a logarithmic leading Regge pole
\begin{eqnarray}\label{eq:asymp_2}
A(s,t)\rightarrow \ln^{\alpha(t)}(s/s_c),
\end{eqnarray}

\noindent respecting the FM bound. 

The very successful model of Donnachie and Landshoff \cite{A.Donnachie.P.V.Landshoff.Phys.Lett.B.296.227.1992}, early proposed by Badatya and Patnaik \cite{R.C.Badatya.P.K.Patnaik.Pramana.15.463-474.1980}, describes the hadronic exchanges remarkably well assuming a simple Regge pole
\begin{eqnarray}\label{eq:dl_1}
A(s,t)= C(t)s^{1.08+0.25t},
\end{eqnarray}

\noindent where $C(t)$ is a constant depending only on $t$. The corresponding $\sigma_{tot}(s)$ of such parameterization is written using the optical theorem as
\begin{eqnarray}\label{eq:sigma_tot_1}
\sigma_{tot}(s)=C(0)s^{0.08},
\end{eqnarray}

\noindent saturating the FM bound. This Regge pole corresponds to one-pole exchange, while the double-pole exchange leads to $\sigma_{tot}(s)\sim \ln(s/s_c)$. The intercept of the Regge pole in this model is defined as a linear function
\begin{eqnarray}\label{eq:intercept_1}
\alpha(t)=\alpha(0)+\alpha' t,
\end{eqnarray}

\noindent where the intercept takes the value $\alpha(0)=\alpha_{\mathbb{P}}=1.08$ and the slope $\alpha'=0.25$ GeV$^{-2}$. 

In the language of physics, this intercept corresponds to a soft pomeron - low momentum transfer - $\alpha_{\mathbb{P}}\approx 1.05\sim1.08$. In contrast, the hard pomeron, predicted to mediate diffractive processes - large momentum transfer - has a value $\alpha_{\mathbb{P}}\approx1.4\sim1.5$. 
This hard pomeron emerges due to the use of the Regge formalism as an analogy to explain the structure-function $F_2(x,Q^2)$ in terms of the Bjorken scale $x$ and the photon virtuality $Q^2$. 

Note that both, the one-pomeron exchange in the Donnachie and Landshoff model or the double-pomeron exchange in the logarithmic Regge pole, leads to intercepts $\gtrsim 1$ \cite{A.Donnachie.P.V.Landshoff.Phys.Lett.B.296.227.1992,S.D.Campos.Phys.Scrip.xx.2020}. However, the one-pomeron exchange with an intercept slightly above 1 violates the FM bound while for the double-pomeron exchange, this violation only occurs for an intercept above 2. If the intercept is equal 2, then the triple-pomeron exchange is favorable, $\sigma_{tot}\approx \ln^2(s/s_c)$. Neither the soft nor the hard pomeron has been discovered yet.


\subsection{Non-subtraction case}

In \cite{S.D.Campos.Phys.Scrip.xx.2020}, one considers only the non-subtraction case. Then, using the optical theorem one has a simple relation for the asymptotic total cross section 
\begin{eqnarray}\label{eq:asymp_3}
\sigma_{tot}(s)\rightarrow \ln^{\alpha_{\mathbb{P}}}(s/s_c).
\end{eqnarray}

In the specified range for $\cos(\theta)$, it respects the FM bound if $\alpha_{\mathbb{P}}\leq 2$, providing a physical relation between the pomeron intercept, $\alpha_{\mathbb{P}}$, and the saturation of this bound. The soft pomeron, if it exists, is the particle allowing the maximum growth to the total cross section, obeying the FM bound. As shall be seen, the phenomenology associated with the $\rho$-parameter is crucial to get the correct pomeron intercept.

\begin{figure}
	\centering{
		\includegraphics[scale=0.37]{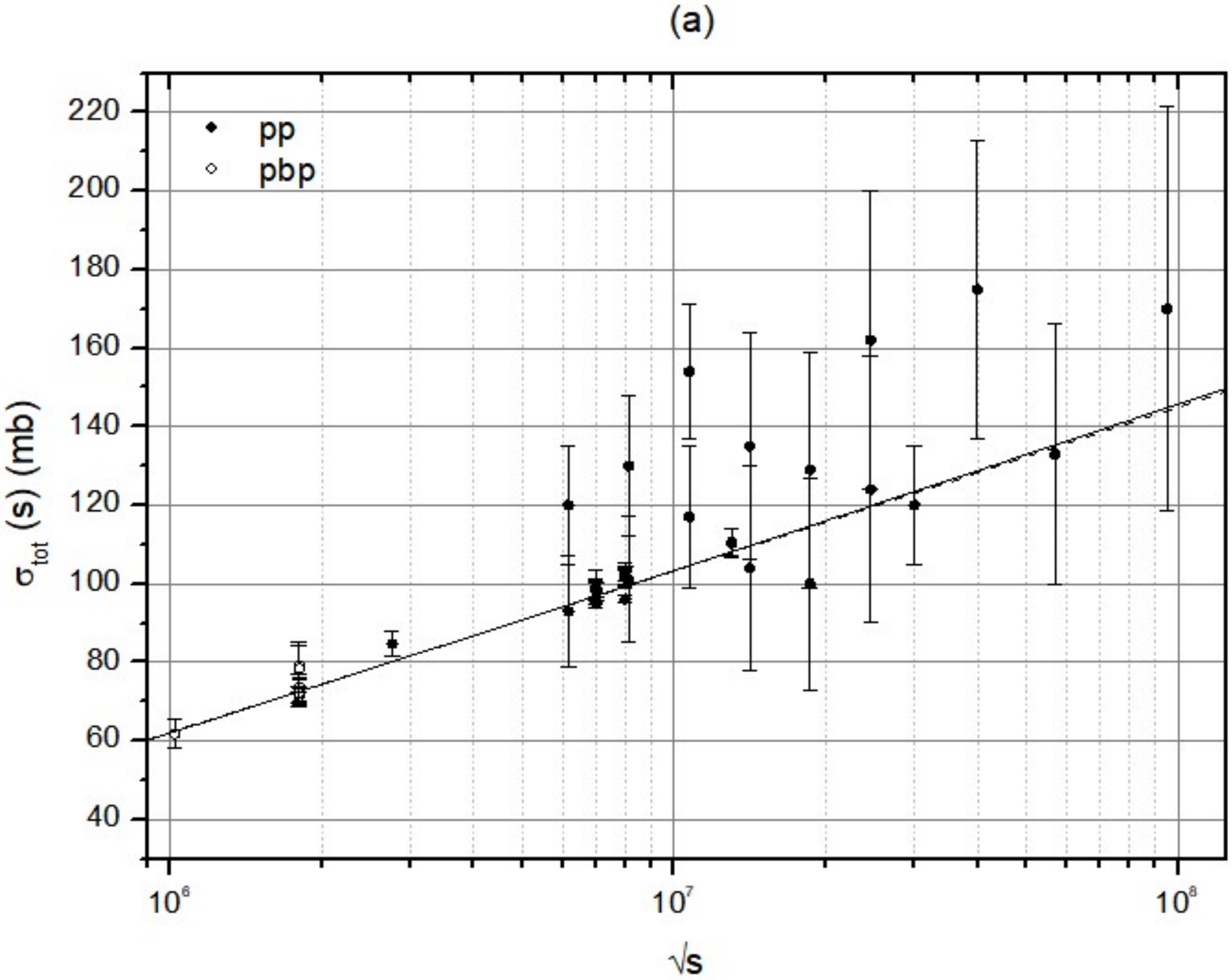}\\
		\includegraphics[scale=0.37]{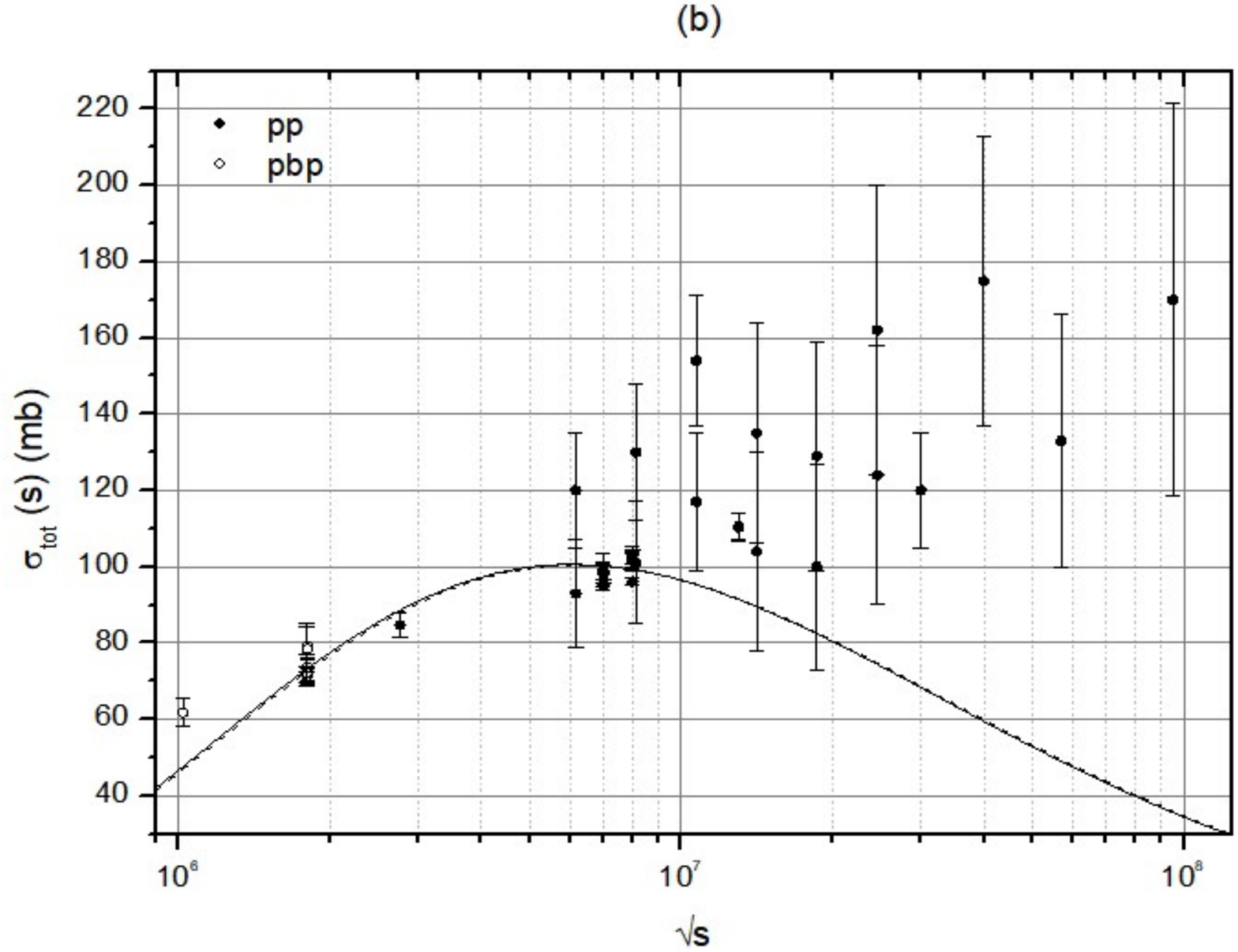} 
		\vspace{-0.5cm}
		\caption{In both panels, the solid line is for the SET 1 and the dashed line is for SET 2. In panel (a) one uses the parameterization (\ref{eq:sig_tot_1}). In panel (b), the parameterization used is given by (\ref{eq:sig_tot_2}). Experimental data are from \cite{PDG-PhysRev-D98-030001-2018,G_Antchev_TOTEM_ Coll_Eur_Phys_J_C79_103_2019}.}
		\label{fig:fig_1}}
\end{figure}

Using a simple parameterization for the total cross section  
\begin{eqnarray}\label{eq:sig_tot_1}
\sigma_{tot}(s)=\beta\ln^{\alpha_{\mathbb{P}}}(s/s_c),
\end{eqnarray}

\noindent where $\beta$ and $\alpha_{\mathbb{P}}$ are free fitting parameters, one can attain the pomeron intercept. Using the SET 1, one obtains the values for the fitting parameters shown in the first line of the Table \ref{tab:table_1}. Figure \ref{fig:fig_1}a shows the curve obtained from the fitting procedure using (\ref{eq:sig_tot_1}). The intercept agrees with a double-pole pomeron exchange. The fitting results using the SET 2 are shown in the second line of the Table \ref{tab:table_1}. From the statistical point of view, the absence of the cosmic-ray data in the SET 2 practically does not alter the results. 

\begin{table}[h!]
	\caption{\label{tab:table_1}Parameters obtained by using (\ref{eq:sig_tot_1}) in the fitting procedures for the SET 1 and 2 and taking $\sqrt{s_c}=25.0$ GeV.}
		\begin{tabular}{c | c | c | c}
			\hline
			SET  ~&~ $\alpha_{\mathbb{P}}$ ~&~ $\beta$ (mb) ~&~$\chi^2/ndf$ \\ 
			\hline
			1    ~&~ 1.05$\pm$0.05 ~&~ 7.54$\pm$0.92 ~&~1.26   \\
			\hline
			2    ~&~ 1.04$\pm$0.05 ~&~ 7.72$\pm$0.98 ~&~1.31  \\
			\hline
			\end{tabular}
\end{table}

\subsection{Subtraction case}

The total cross section with one subtraction can be written using the following normalization of the optical theorem
\begin{eqnarray}\label{eq:sub_1}
s\sigma_{tot}(s)=\mathrm{Im}A(s).
\end{eqnarray}

This normalization implies in the logarithmic leading Regge pole
\begin{eqnarray}\label{eq:sub_2}
\sigma_{tot}(s)\approx \frac{\ln^{\alpha_{\mathbb{P}}}(s/s_c)}{s}.
\end{eqnarray}

First of all, one notices that to tame the fast decreasing of the total cross section entailed by the subtraction, it is necessary a $\alpha_{\mathbb{P}}$ far above from the expected saturation of the FM bound, $\alpha_{\mathbb{P}}\rightarrow 2$. Indeed, to give some physical meaning for $\alpha_{\mathbb{P}}$ is now a hard task. The effect of using $s$ in the above result can be observed by adopting the parameterization given below
\begin{eqnarray}\label{eq:sig_tot_2}
\sigma_{tot}(s)=\beta\frac{\ln^{\alpha_{\mathbb{P}}}(s/s_c)}{s}.
\end{eqnarray}

The fitting results obtained by using the parameterization (\ref{eq:sig_tot_2}) to the SET 1 and 2 are shown in the Table \ref{tab:table_2}. Of course, the subtracted case cannot be used as a realistic parameterization to describe the $\sigma_{tot}(s)$. Figure \ref{fig:fig_1}b shows the curve obtained from the fitting procedure using (\ref{eq:sig_tot_2}).  

\begin{table}[h!]
	\caption{\label{tab:table_2}Fitting parameters obtained by using (\ref{eq:sig_tot_2}) in the fitting procedure to the SET 1 and 2 and taking $\sqrt{s_c}=25.0$ GeV.}
		\begin{tabular}{c | c | c | c}
			\hline
			SET  ~&~ $\alpha_{\mathbb{P}}$ ~&~ $\beta$ (mb) ~&~$\chi^2/ndf$ \\ 
			\hline 
			1 ~&~  11.02$\pm$0.09 ~&~ 1.99$\times 10^{-5}\pm 4.9\times 10^{-6}$ ~&~ 5.14 \\
			\hline
			2 ~&~  10.99$\pm$0.10 ~&~ 2.15$\times 10^{-5}\pm 5.7\times 10^{-6}$ ~&~ 5.89 \\
			\hline
		\end{tabular}
\end{table}

However, if we release the subtraction mechanism by introducing the $\delta$- index as a measurement of the deviation of the non-subtraction to the subtraction case, then one writes the parameterization
\begin{eqnarray}\label{eq:sub_3}
\sigma_{tot}(s)= \beta\frac{\ln^{\alpha_{\mathbb{P}}}(s/s_c)}{s^\delta}.
\end{eqnarray}

\begin{figure}
	\centering{
		\includegraphics[scale=0.37]{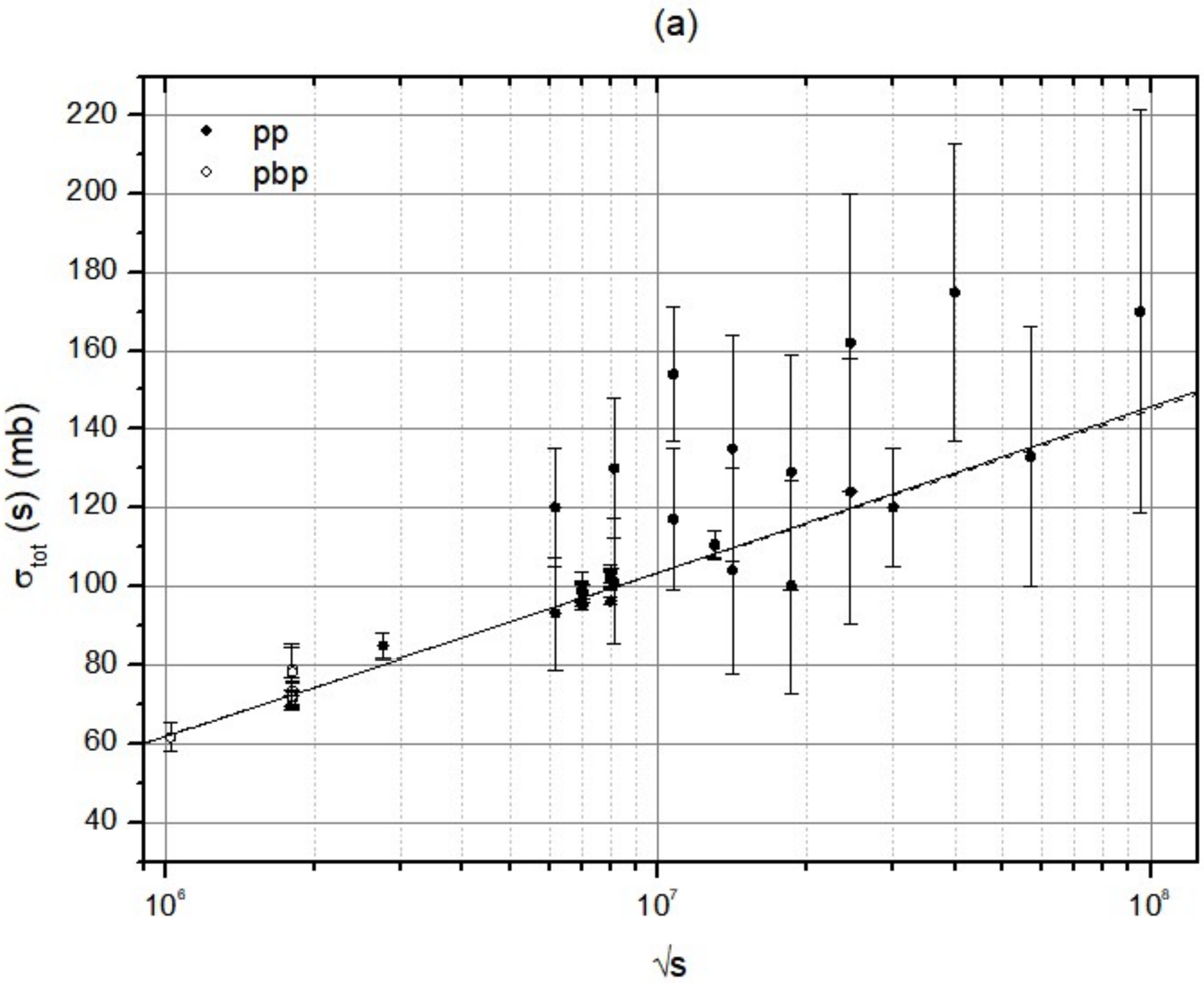}\\
		\includegraphics[scale=0.37]{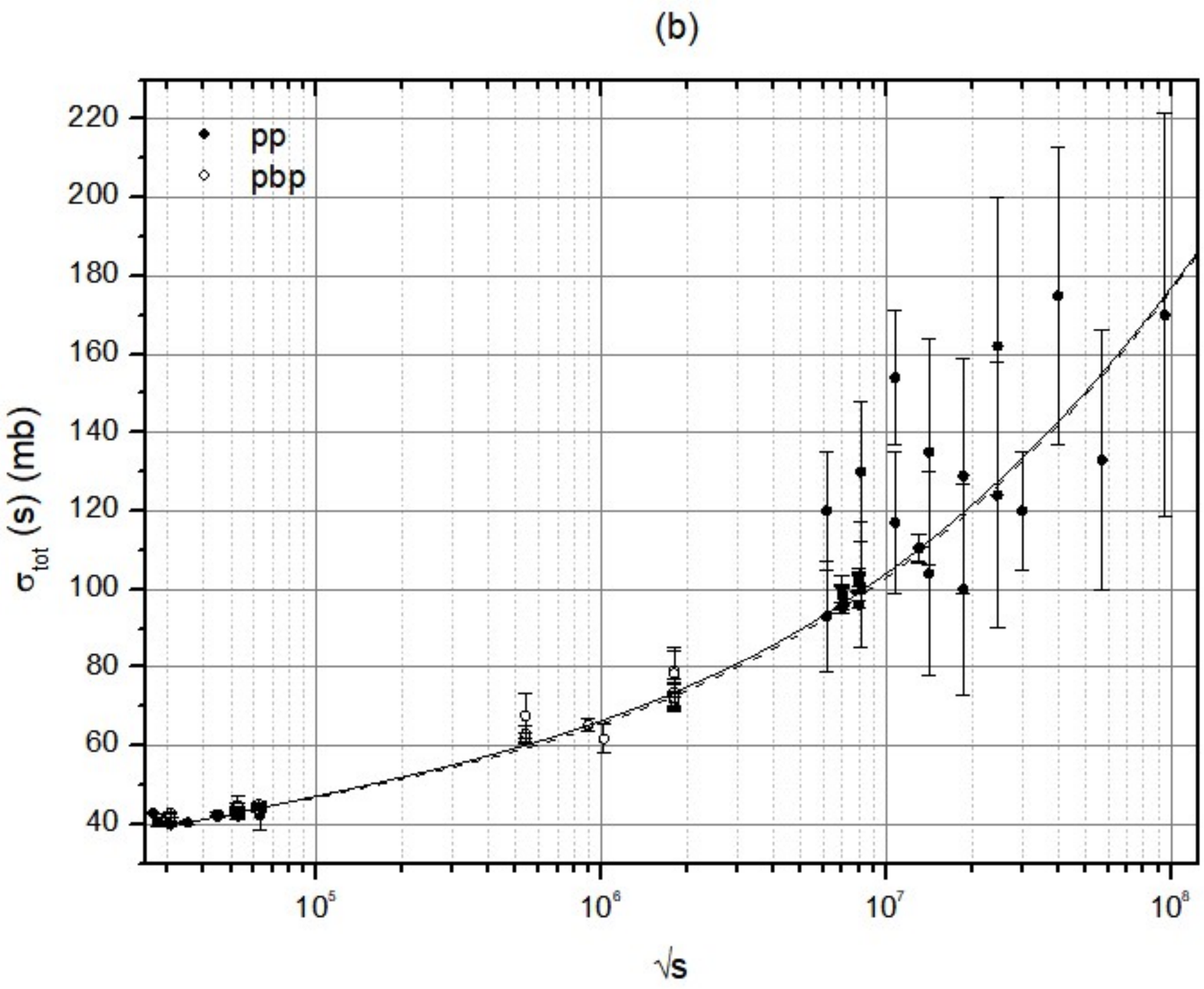} 
	    \vspace{-0.5cm}
		\caption{In both panels, the solid line is for the SET 1 and the dashed line is for SET 2. In panel (a) one uses the parameterization (\ref{eq:sub_3}). In panel (b), the parameterization used is given by (\ref{eq:tcs_2}). Experimental data are from \cite{PDG-PhysRev-D98-030001-2018,G_Antchev_TOTEM_ Coll_Eur_Phys_J_C79_103_2019}.}
		\label{fig:fig_2}}
\end{figure}

Using the SET 1 and 2, one obtains very small values for the $\delta$-index (near zero). In Table \ref{tab:table_3} are displayed the fitting parameters. The $\delta$-index introduces an error in the fitting parameters higher than the non-subtraction case. However, the central value of each parameter is practically the same. The fitting result is shown in Figure \ref{fig:fig_2}a.

\begin{table}[h!]
	\caption{\label{tab:table_3}Fitting parameters obtained by using (\ref{eq:sub_3}) in the fitting procedure to the SET 1 and 2 and taking $\sqrt{s_c}=25.0$ GeV.}
		\begin{tabular}{c | c | c | c |c}
			\hline
			SET  ~&~ $\alpha_{\mathbb{P}}$ ~&~ $\beta$ (mb) ~&~$\delta$ ~&~ $\chi^2/ndf$ \\ 
			\hline 
			1 ~&~  1.05$\pm$0.82 ~&~ 7.55$\pm$8.07  ~&~ 0.00$\pm$0.08 ~&~ 1.29 \\
			\hline
			2 ~&~  1.04$\pm$1.07 ~&~ 7.72$\pm$10.70 ~&~ 0.00$\pm$0.11 ~&~ 1.77 \\
			\hline
		\end{tabular}
\end{table}

To circumvent the need for the use of the $\delta$-index, one introduces the approximation \cite{S.D.Campos.Phys.Scrip.xx.2020}
\begin{eqnarray}\label{eq:app_1}
\frac{1}{(s/s_c)^{\delta}}\approx \frac{1}{a(\ln^{\epsilon}(s/s_c))}, 
\end{eqnarray}

\noindent which is a consequence of the fact that the inequality given below holds for $0<s_c\leq s$ and $0\leq\epsilon\leq\delta\in \mathbb{R}$
\begin{eqnarray}\label{eq:app_2} 
\ln^\epsilon(s/s_c)\leq (s/s_c)^\delta\Rightarrow\frac{1}{(s/s_c)^\delta}\leq\frac{1}{\ln^\epsilon(s/s_c)}.
\end{eqnarray}

In particular, for some $0<a\in\mathbb{R}$, the approximation (\ref{eq:app_1}) can be used, resulting in the total cross section
\begin{eqnarray}\label{eq:sub_4}
\sigma_{tot}(s)\approx \beta'\ln^{\alpha_{\mathbb{P}}'}(s/s_c),
\end{eqnarray}

\noindent where $\alpha_{\mathbb{P}}'=\alpha_{\mathbb{P}}-\epsilon$ and $\beta'=c/a$. The choice of $a$ is not unique, and in general, its value depends on the energy range where the experimental data are being analyzed. However, in the fitting procedure, it is absorbed by $\beta'$. Therefore, the expression (\ref{eq:sub_4}) corresponds to the subtraction case and it is exactly equal to the non-subtraction one given by (\ref{eq:sig_tot_1}).

\subsection{The cut case}

As well-known, the singularities play a fundamental role in the determination of the divergences of the partial-wave amplitude. Neglecting the signature of the partial-wave amplitude, one may write the singularities as
\begin{eqnarray}\label{eq:sing_1}
A_l(t)\propto\int_z^{\infty}\mathrm{disc}A(s,t)Q_l(z')dz',
\end{eqnarray}

\noindent where $Q_l(z')$ is the Legendre function of the second kind and $\mathrm{disc}A(s,t)$ is the discontinuity across the $l$-plane cut. The easy way to obtain singularities from (\ref{eq:sing_1}) is to associate the $\mathrm{disc}A(s,t)$ with the Legendre function of first kind, $P_{\alpha_c(t)}(z)$, where $\alpha_c(t)$ is the cut. Then, in the simplest case,
\begin{eqnarray}
\mathrm{disc}A(s,t)\propto P_{\alpha_c(t)}(z).
\end{eqnarray}

In this picture, one uses the properties of the Legendre functions \cite{erdelyi_book}
\begin{eqnarray}\label{eq:leg_1}
\int_1^{\infty}P_{\alpha_c(t)}(z)Q_l(z)dz=\frac{1}{(l+1+\alpha_c(t))(l-\alpha_c(t))},
\end{eqnarray}

\noindent revealing a pole for $l=\alpha_c(t)$ ($l\in\mathbb{R}$). 

The Watson-Sommerfeld representation for the partial-wave expansion in the complex angular plane can be written as
\begin{eqnarray}\label{eq:ws_1}
A(s,t)_{s\rightarrow\infty}\propto R_p(s,t)+R_c(s,t)+\mathrm{vanishing}~\mathrm{terms},
\end{eqnarray}

\noindent where $R_p(s,t)$ and $R_c(s,t)$ stands for the Regge pole and the Regge cut contributions for the scattering amplitude. The logarithmic Regge pole introduced in \cite{S.D.Campos.Phys.Scrip.xx.2020} gives for the first term in the r.h.s. of (\ref{eq:ws_1})
\begin{eqnarray}\label{eq:reggepole}
R_p(s,t)\approx P_{\alpha(t)}(\cos(\theta))_{s\rightarrow\infty}^{s>\!>|t|}\longrightarrow \frac{\mathrm{\Gamma}(2l+1)}{\mathrm{\Gamma}^2(l+1)}\ln^{\alpha(t)}(s/s_c)\approx \ln^{\alpha(t)}(s/s_c).
\end{eqnarray}

If a branch point with a cut is encountered during the deformation of the contour in the complex momentum angular plane, then it contributes to the asymptotic behavior of the scattering amplitude. Then, one should take into account the cut contribution. One writes
\begin{eqnarray}\label{eq:cut_1}
R_c(s,t)\propto-\frac{1}{i}\int^{\alpha_{c}(t)}dl(2l+1)\mathrm{disc}A(l,t)\frac{P_l(-z)}{\sin\pi l}.
\end{eqnarray}

As stated above, the functional form of the discontinuity is not known \textit{a priori}, and there is no phenomenological approach for it. Therefore, the only way to go through this point is by using an ansatz as performed in Ref. \cite{P.D.B.Collins.book.1977}.

Using the property $\mathrm{Im}P_l(z)=-P_l(-z)\sin\pi l$, $z<-1$, in the logarithmic Regge approach, one adopts $\mathrm{disc}A(l,t)=(\alpha_c(t)-l)^{1+\beta(t)}$, obtaining the approximation
\begin{eqnarray}\label{eq:cut_2}
R_c(s,t)\approx-\ln^{\alpha_c(t)}(s/s_c)(\ln\ln(s/s_c))^{-(2+\beta(t))}
\end{eqnarray}

\noindent where $-2<\beta(t)$ is a function of $t$ only. Taking into account the results (\ref{eq:reggepole}) and (\ref{eq:cut_2}), then the asymptotic scattering amplitude is written as
\begin{eqnarray}\label{eq:final}
A(s,t)\approx \ln^{\alpha(t)}(s/s_c) - \ln^{\alpha_c(t)}(s/s_c)(\ln\ln(s/s_c))^{-(2+\beta(t))}.
\end{eqnarray} 

Of course, the above result is strongly dependent on the form of $\mathrm{disc}A(l,t)$ and there is no information about $\beta(t)$. The Regge cut $\alpha_c(t)$ is also an unknown function of $t$. One suppose here, that $\alpha_c(t)$ and $\beta(t)$ are a real-valued functions as well as they assume finite values at $t=0$. 

It is important to stress that $\sigma_{tot}(s)$, independently of the functional form of $\mathrm{disc}A(l,t)$, should obey the FM bound. Then, the rise of (\ref{eq:final}) is controlled by $\ln^2(s/s_c)$. It is not difficult to see that (hereafter, $\alpha_c(0)=\alpha_c$)
\begin{eqnarray}
0\leq 1-\ln^{\alpha_c-\alpha_{\mathbb{P}}}(s/s_c)(\ln\ln(s/s_c))^{-(2+\beta(0))}
\end{eqnarray}

\noindent implies that in the asymptotic regime $s\rightarrow\infty$
\begin{eqnarray}\label{eq:inel}
\frac{\alpha_c-\alpha_{\mathbb{P}}}{2+\beta(0)}\leq 0 \Rightarrow \alpha_c\leq \alpha_{\mathbb{P}}
\end{eqnarray}

\noindent for $\beta(0)>-2$. The inequality $\alpha_c\leq \alpha_{\mathbb{P}}$ implies the Regge cut is bounded by the Regge pole.  According to the general belief, the subleading (secondary) contributions are important below 1.0 TeV. Then, at ISR energies the effect of a secondary and leading contributions are mixed, while at LHC energies the leading contribution dominates.  Therefore, the Regge cut (\ref{eq:final}) may be used to explain the experimental data behavior from the minimum value of the total cross section up to $\sim$ 1.0 TeV since the role of the cut is now clear: it represents the mixed contributions below the logarithmic Regge pole dominance according to inequality (\ref{eq:inel}).

\begin{table}[h!]
	\caption{\label{tab:table_4}Fitting parameters obtained by using (\ref{eq:tcs_2}) in the fitting procedure to the SET 3 and 4 and taking $\sqrt{s_c}=15.0$ GeV.}
		\small{\begin{tabular}{c | c | c | c | c | c | c  }
			\hline
      SET & $\beta_1$ (mb) & $\alpha_{\mathbb{P}}$ & $\beta_2$ (mb)  &    $\alpha_c$ & $\beta(0)$      &  $\chi^2/ndf$ \\
			\hline	
3   & 0.01$\pm$0.01  & 3.22$\pm$0.43         & -31.32$\pm$1.10 & 0.32$\pm$0.04 & -1.86$\pm$0.02  &  2.91    \\
			\hline
4   & 0.01$\pm$0.01  & 3.24$\pm$0.50         & -31.29$\pm$1.19 & 0.32$\pm$0.04 & -1.86$\pm$0.02  &  3.36  \\
			\hline
		\end{tabular}}
\end{table}

To take into account the Regge pole and the Regge cut contributions for the total cross section, one uses the simple parameterization
\begin{eqnarray}\label{eq:tcs_2}
\sigma_{tot}(s)=\beta_1 \ln^{\alpha_{\mathbb{P}}}(s/s_c)-\beta_2\ln^{\alpha_c}(s/s_c)(\ln\ln(s/s_c))^{-(2+\beta(0))},
\end{eqnarray}

\noindent to fit the SET 3 and 4. The fitting results are displayed in Table \ref{tab:table_2}. The Figure \ref{fig:fig_2}b shows the curve obtained for the parameterization (\ref{eq:tcs_2}). 

It is important to stress that there is no constraint on the fitting parameters. Then, one allows the pomeron intercept to assume any value necessary to fit the experimental data. The result, at first glance, shows a pomeron intercept far above from the saturation of the FM bound - a supercritical value. Then, a control mechanism should be imposed to tame the growth of the total cross section as $s\rightarrow\infty$. This control mechanism, as shall be seen, is obtained by looking at the $\rho$-parameter experimental data.   




\section{The $\rho$-parameter and slope B(s,t)}\label{sec:dr}

As well-known, the validity of the Cauchy theorem is crucial for the convergence of the integral dispersion relation. This kind of relationship allows the obtaining of the real part of the forward scattering amplitude from the imaginary part. First of all, one writes the scattering amplitude as the sum of the even (+) and odd (-) amplitudes as
\begin{eqnarray}\label{eq:scatt}
A(s,t)=A_+(s,t)\pm A_-(s,t),
\end{eqnarray}

\noindent where $A_\pm(s,t)=\mathrm{Re}A_\pm(s,t)+i\mathrm{Im}A_\pm(s,t)$ are the crossing even (+) and odd (-) amplitudes. The integral form of the dispersion relations is a consequence of analyticity, unitarity, and crossing properties. In the non-subtraction case, it is simply written as ($t=0$)
\begin{eqnarray}\label{eq:dr_1}
\mathrm{Re}A_+(s)=\frac{2s}{i}P\int_{s_{min}}^\infty ds' \frac{\mathrm{Im}A_+(s')}{s-s'},
\end{eqnarray}

\noindent where $P$ is the principal value of Cauchy integral. The convergence of the above integral can be ensured by using the subtraction procedure, i.e. by rewritten the scattering amplitude as
\begin{eqnarray}\label{eq:scatt_2}
\tilde{A}(s)=\left|\frac{A(s)}{s}\right|
\end{eqnarray}

\noindent which result in the subtraction term 
\begin{eqnarray}
\mathrm{Re}\tilde{A}_+(s)=K+\frac{2s^2}{i}P\int_{s_{min}}^\infty ds' \frac{\mathrm{Im}\tilde{A}_+(s')}{s(s^2-s'^2)},
\end{eqnarray}

\begin{eqnarray}
\mathrm{Re}\tilde{A}_-(s)=\frac{2s^2}{i}P\int_{s_{min}}^\infty ds' \frac{\mathrm{Im}\tilde{A}_-(s')}{(s^2-s'^2)},
\end{eqnarray}

\noindent where $K$ is the subtraction constant. Of course, this method is valid only for a finite number of subtractions \cite{Y.S.Jin.A.Martin.Phys.Rev.135.B1375.1964}. However, the integral dispersion relations are very restrictive, since to know the value of the real part to a specific value one should know the value of the imaginary part in the whole plane. Then, despite its rigorous formulation, the use of integral dispersion relations is of little interest in the Regge theory.

On the other hand, the derivative dispersion relation can be used in the present case \cite{J.B.Bronzan.G.L.Kane.U.P.Sukhatme.Phys.Lett.B49.272.1974,K.Kang.B.Nicolescu.Phys.Rev.D11.2461.1975}. These derivative relations can be written in the first-order approximation for the odd and even amplitudes as \cite{S.D.Campos.EPJC.47.171.2006}
\begin{eqnarray}
\frac{\mathrm{Re}A_+(s,t)}{s}=\frac{K}{s}+\frac{\pi}{2}\frac{d}{d\ln s}\frac{\mathrm{Im}A_+(s)}{s},
\end{eqnarray}

\begin{eqnarray}
\frac{\mathrm{Re}A_-(s,t)}{s}=\frac{\pi}{2}\left(1+\frac{d}{d\ln s}\right)\frac{\mathrm{Im}A_-(s)}{s}.
\end{eqnarray}

Considering (\ref{eq:sig_tot_1}) and (\ref{eq:tcs_2}), it is possible to obtain real part of the scattering amplitude in the subtraction as well as in the Regge cut. Then, the real part of the scattering amplitude can be used to define the $\rho$-parameter  
\begin{eqnarray}
\rho(s)=\frac{\mathrm{Re}A(s)}{\mathrm{Im}A(s)}.
\end{eqnarray}

\begin{figure}
	\centering{
		\includegraphics[scale=0.40]{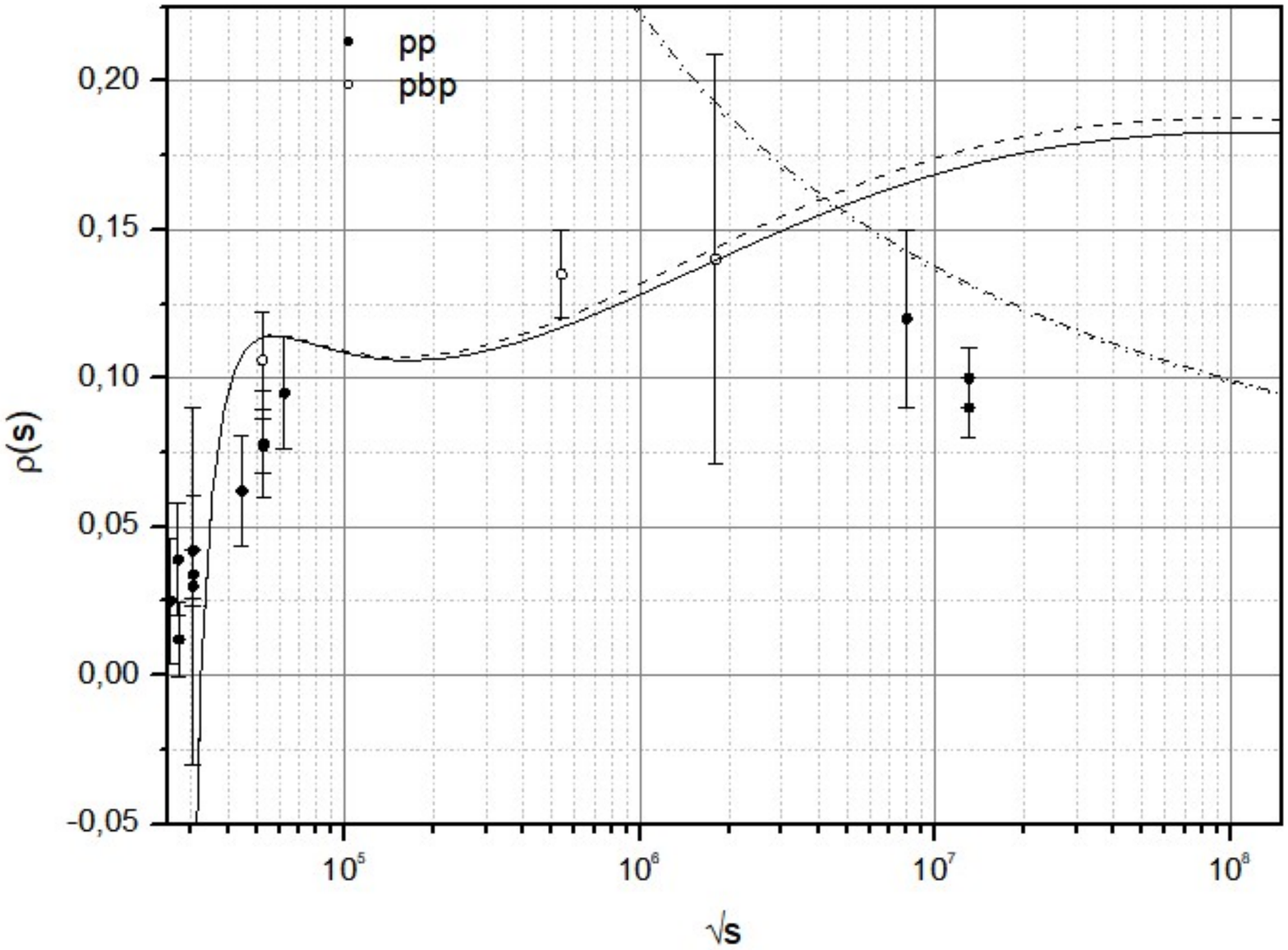}
		\vspace{-0.5cm}
		\caption{The dot-dashed and dotted lines are obtained applying (\ref{eq:sig_tot_1}) to the SET 1 and 2, respectively. The solid and dashed lines are obtained using (\ref{eq:tcs_2}) to the SET 3 and 4, respectively. Experimental data are from \cite{PDG-PhysRev-D98-030001-2018,G_Antchev_TOTEM_ Coll_Eur_Phys_J_C79_103_2019}.}
		\label{fig:fig_3}}
\end{figure}

Without loss of generality, one set $K=0$, since the influence of this parameter is restricted to the low energy region, and the main interest here is to understand the asymptotic regime. Then, all parameters were previously obtained from the fits to the total cross section. 

In Figure \ref{fig:fig_3} one has the predictions for the $\rho$-parameter. The results from the logarithmic Regge pole (\ref{eq:sig_tot_1}) are represented by the dotted line and by the dot-dashed line, SET 1 and 2, respectively. These curves represent the pure contribution coming from the double pomeron exchange. Therefore, these results are not able to reproduce the low energy behavior of the $\rho$-parameter. Using the parameterization (\ref{eq:tcs_2}), the solid and dashed lines, respectively, display the Regge cut contribution obtained from the fitting procedures for the SET 3 and 4. These predictions show behavior in the high energy regime not present in the experimental data. There is a fast-rising introduced by the experimental data below $\sqrt{s}=1.0$ TeV, which seems not present in the experimental data above this energy since the fittings for the SET 1 and 2 shown a pomeron intercept $\alpha_{\mathbb{P}}\approx 1.05$.

An attempt to solve this problem can be done by using the recent experimental data for the $\rho$-parameter at $\sqrt{s}=13.0$ TeV. These experimental data suggest a double pomeron intercept taming the rise of the total cross section. Then, taking into account the above discussion, one introduces the constraint $\alpha_{\mathbb{P}}\leq 1$ to reduce the fast-rising of $\sigma_{tot}(s)$ above 1.0 TeV, introduced by (\ref{eq:tcs_2}). The Table \ref{tab:table_5} shows the fitting parameters only for the SET 3. Despite the high value to $\chi^2/ndf$, the fitting parameters seems to be able to reproduce the growth of $\sigma_{tot}(s)$ as $s\rightarrow\infty$.

\begin{table}[h!]
	\caption{\label{tab:table_5}Fitting parameters obtained by using (\ref{eq:tcs_2}) in the fitting procedure to the SET 3, taking $\sqrt{s_c}=15.0$ GeV and with the constraint $\alpha_{\mathbb{P}}\leq 1$.}
		\small{\begin{tabular}{c | c | c | c | c | c}
			\hline
$\beta_1$ (mb) & $\alpha_{\mathbb{P}}$ & $\beta_2$ &$\alpha_c(0)$ &$\beta(0)$ &$\chi^2/ndf$   \\                                   
			\hline
6.12$\pm$5.59 & 1.00$\pm$0.27 & -30.96$\pm$3.29 & -0.14$\pm$0.31 & -1.92$\pm$0.04 & 4.36 \\
		
			\hline
			
		\end{tabular}}
\end{table}

\begin{figure}
	\centering{
		\includegraphics[scale=0.37]{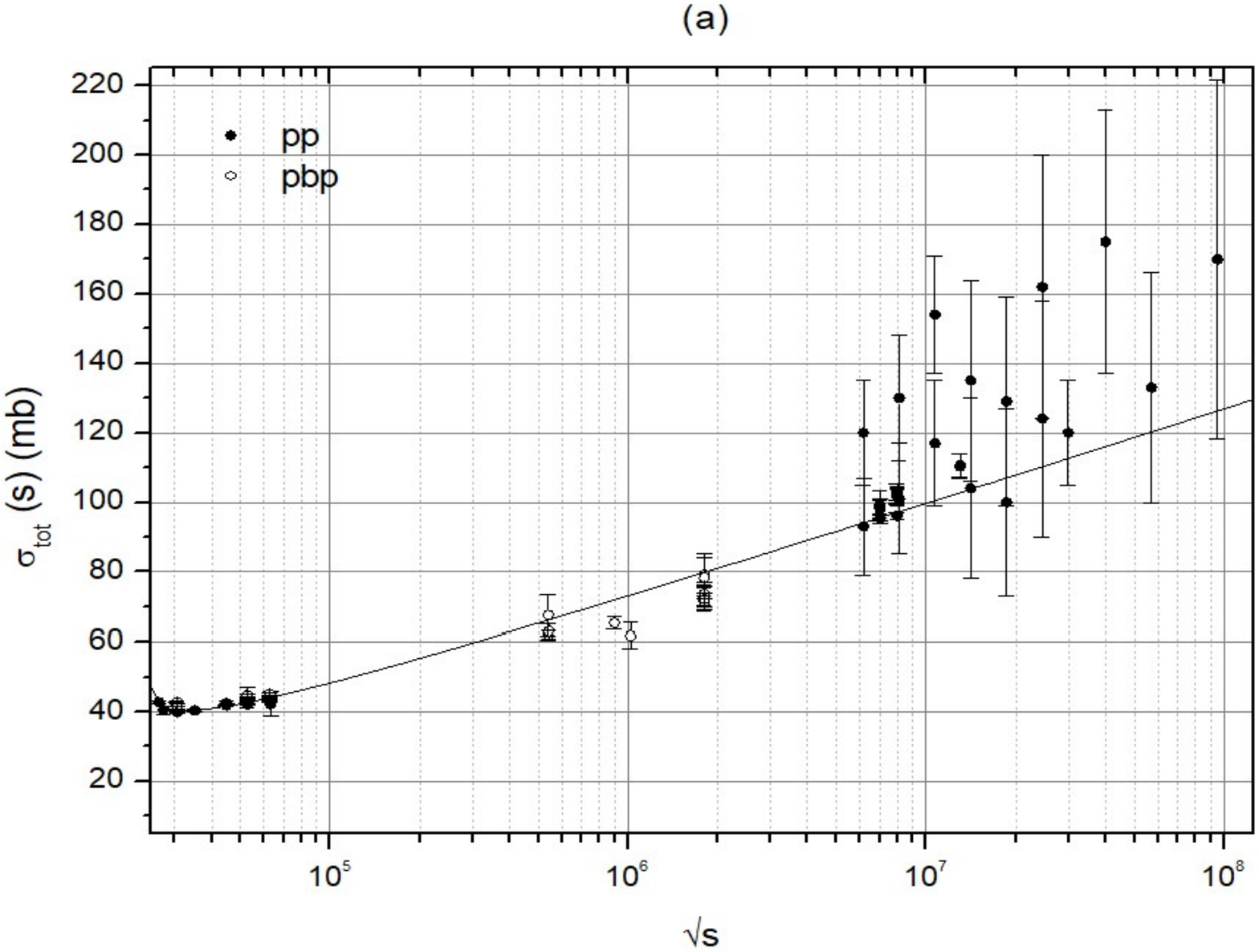}\\
		\includegraphics[scale=0.37]{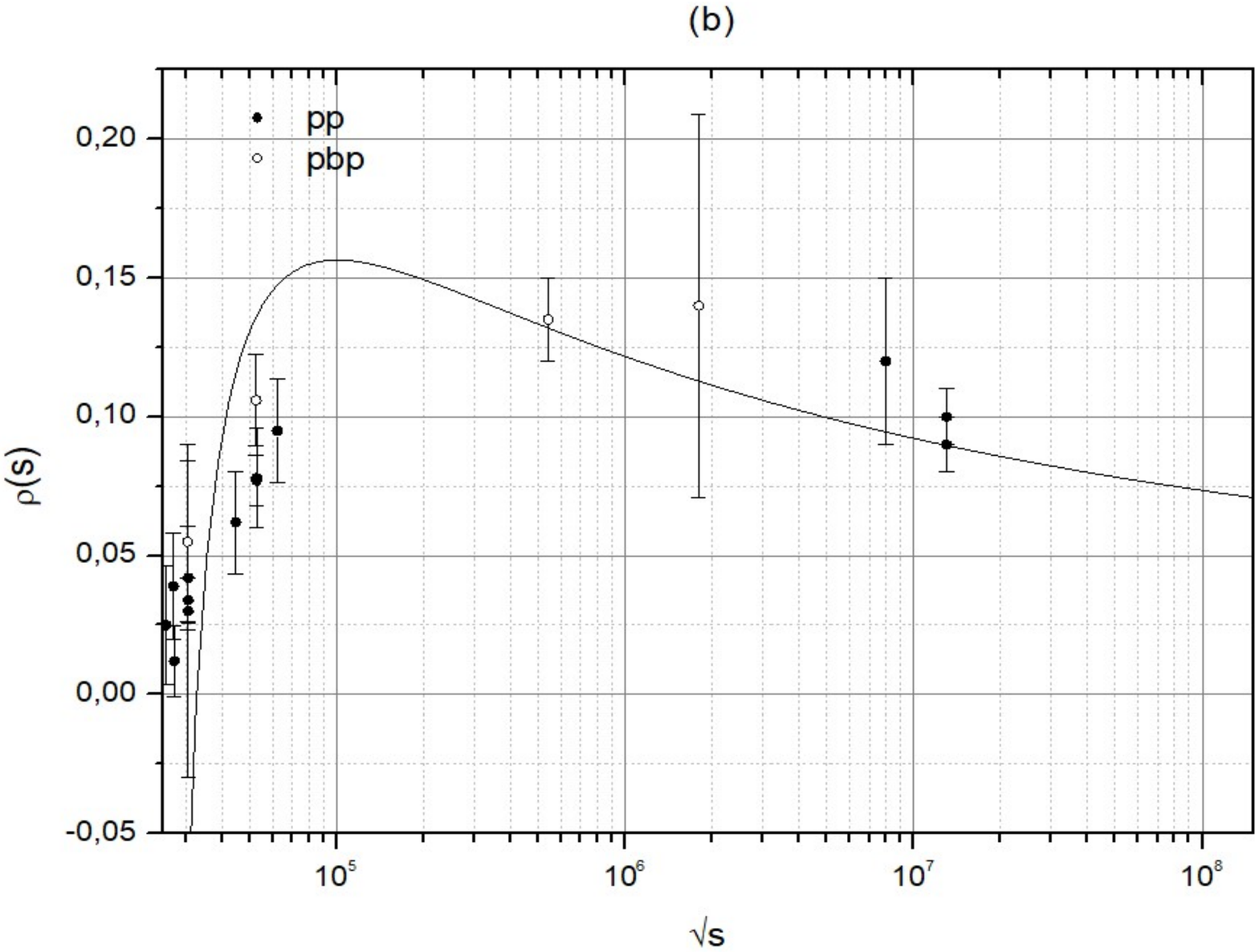} 
		\vspace{-0.5cm}
		\caption{The experimental data are from SET 3 in panel (a), and the parameterization is given by (\ref{eq:tcs_2}) with the pomeron constraint $\alpha_{\mathbb{P}}\leq 1$. In panel (b), one presents the prediction for the $\rho$-parameter under the double-pomeron constraint.}
		\label{fig:fig_4}}
\end{figure}

The Figure \ref{fig:fig_4}a shows the fitting results for $\sigma_{tot}(s)$ using the parameterization (\ref{eq:tcs_2}) under the double-pomeron constraint. In Figure \ref{fig:fig_4}b, one shows the prediction for the $\rho$-parameter. Despite this naive parameterization, one can observe that the only way to reproduce the high energy behavior of $\rho$ at $\sqrt{s}=13.0$ TeV is assuming a double pomeron exchange in the logarithmic Regge pole representation.

A possible understanding here is: the fast rise of $\sigma_{tot}(s)$ in the logarithmic Regge approach cannot be analyzed independently of the $\rho$-parameter. In particular, the $\rho$ value at $\sqrt{s}=13.0$ TeV is crucial to determine an upper bound to the pomeron intercept. It should be stressed that it is a feature of the Regge theory: the phenomenology. 

Another physical quantity that can be predicted from the fitting procedures performed here is the (forward) slope of the elastic differential cross section $d\sigma/dt$, defined as
\begin{eqnarray}
B(s,t\rightarrow 0)=\frac{d}{dt}\left(\ln{\frac{d\sigma}{dt}}\right)_{t=0}.
\end{eqnarray}

In the present approach, using the simple asymptotic parameterization (\ref{eq:sig_tot_1}), for example, one obtains for the forward case
\begin{eqnarray}\label{eq:slope}
B(s)=B_0+2\alpha'\ln(\ln s/s_0),
\end{eqnarray}

\noindent where $\sqrt{s_0}$ is some initial energy not necessarily related with $\sqrt{s_c}$. The value $s_0$ can be taken from some current known experimental result for the slope, being used as input to predict the slope at $s$. For example, using the TOTEM result $B_0=19.9\pm0.3$ GeV$^{-2}$ at $\sqrt{s_0}=7.0$ TeV \cite{G.Antchev.etal.TOTEM.Coll.Europhys.Lett.101.21002.2013} and $\alpha'=0.25$ GeV$^{-2}$, one has the prediction for the elastic slope $B(s)=20.0\pm0.3$ GeV$^{-2}$ at $\sqrt{s}=13.0$ TeV. As well-known, the LHC result at $\sqrt{s}=13.0$ TeV is $B(s)=20.36\pm0.19$ GeV$^{-2}$ \cite{G_Antchev_TOTEM_ Coll_Eur_Phys_J_C79_103_2019}, in accordance with the value predicted here. On the other hand, starting from the TOTEM result, one cannot reproduce, for example, the unexpected value for the slope encountered by E710 Collaboration at $\sqrt{s}=1.8$ TeV, $B(s)=16.98\pm0.25$ GeV$^{-2}$ \cite{N.A.Amos.etal.E710.Coll.Phys.Lett.B247.124.1990}. 

The prediction capability of (\ref{eq:slope}) is, of course, limited by the constraint $\sqrt{s}\geq\sqrt{e s_0}$. This implies, for example, that the next energy above the LHC at $\sqrt{s}=13.0$ TeV that one can predict the slope is at $\sqrt{s}\approx 22$ TeV (the same for the $\rho$-parameter). Therefore, using the LHC result as the initial value $\sqrt{s_0}=13.0$ TeV, one obtains $B(s)=20.38$ GeV$^{-2}$ at $\sqrt{s}=22.0$ TeV, indicating a very slow increase for the slope. At the same energy, the slope from the usual leading Regge pole is $B(s)=20.88$ GeV$^{-2}$.

On the other hand, if instead (\ref{eq:sig_tot_1}) one uses the parameterization considering the logarithmic Regge cut (\ref{eq:tcs_2}), then one has for the slope in the forward case
\begin{eqnarray}
B(s)=B_0+3[(\alpha'+\alpha_c'(0))\ln(\ln s/s_0)-\beta'(0)\ln(\ln(\ln s/s_0))].
\end{eqnarray}

Of course, the importance of the factor $\ln(\ln(\ln s/s_0))$ in the r.h.s. of the above result depends strongly on the values of $s$ and $s_0$ used. For example, from the E710 to the TOTEM energy, one obtain $\ln(\ln(\ln s/s_0))=-7.49\times 10^{-4}$ while from the TOTEM to the LHC, $\ln(\ln(\ln s/s_0))=-1.54$. Therefore, one sets, for the sake of simplicity, $\beta'(0)=0$. Then, one writes for the slope
\begin{eqnarray}\label{eq:slope_linear}
B(s)\approx B_0+3(\alpha'+\alpha_c'(0))\ln(\ln s/s_0).
\end{eqnarray}

The parameter $\alpha_c'(0)$ is unknown and strongly influences the slope. For example, using the TOTEM result $B_0=19.9\pm0.3$ GeV$^{-2}$ at $\sqrt{s_0}=7.0$ TeV and taking $\alpha_c'(0)=0.25$ GeV$^{-2}$, then one has for the LHC result at $\sqrt{s}=13.0$ TeV, $B(s)=20.22\pm0.3$ GeV$^{-2}$. Furthermore, using the E710 experimental result and $\alpha_c'(0)=0.25$ GeV$^{-2}$, one has $B(s)=18.47\pm 0.25$ GeV$^{-2}$ for the TOTEM at $\sqrt{s}=7.0$ TeV. 

However, setting $\alpha_c'(0)=0.50$ GeV$^{-2}$ and starting from the TOTEM energy, then for the LHC at $\sqrt{s}=13.0$ TeV, one gets $B(s)=20.38\pm0.3$ GeV$^{-2}$. Nonetheless, using the E710 at $\sqrt{s}=1.8$ TeV and the TOTEM at $\sqrt{s}=7.0$ TeV, then the slope is $B(s)=19.22\pm0.25$ GeV$^{-2}$. The last result is better than using $\alpha_c'(0)=0.25$ GeV$^{-2}$, but still below the TOTEM result. 

On the other hand, assuming $B_0$ and $\alpha_c'(0)$ as free parameters, and setting $\alpha'=0.25$ GeV$^{-2}$ and $\sqrt{s_0}=15.0$ GeV, then, one can fit the experimental data for the slope above 1.0 TeV ($pp$ and $p\bar{p}$). In this energy range, the slope seems to be a linear function of $s$ and, therefore, one can use the result (\ref{eq:slope_linear}) to fit these experimental data \cite{PDG-PhysRev-D98-030001-2018}. Thus, one obtains $\alpha_c'(0)=2.91\pm 0.25$ GeV$^{-2}$ and $B_0=-3.51\pm 1.73$ GeV$^{-2}$.

It is interesting to note that it is a very hard task to access the slope $B(s)$ since it is not extracted at $t=0$. In general, the slope is obtained considering very small ranges of $t\approx 0$, which implies Coulomb and Nuclear contributions to the scattering amplitude. For example, for the E710 Collaboration \cite{N.A.Amos.etal.E710.Coll.Phys.Lett.B247.124.1990}, the interval for $t$ is $[0.04, 0.29]$ GeV$^2$ and for the TOTEM Collaboration one has $t\in[0.012, 0.2]$ GeV$^2$ \cite{G_Antchev_TOTEM_ Coll_Eur_Phys_J_C79_103_2019}. Then, these experimental values may carry some dependence on the momentum transfer. 

\section{Conclusions}\label{sec:critical}

One obtains here the leading Regge pole with one subtraction as well as the Regge cut, both in the logarithmic representation introduced previously in \cite{S.D.Campos.Phys.Scrip.xx.2020}. The fitting procedures for the subtraction case led to unrealistic results to $\sigma_{tot}(s)$, i.e. to a decreasing faster than the rise of the experimental data, turning the subtraction a problem in the logarithmic Regge pole. 

Trying to understand this problem, one introduces a small parameter, the $\delta$-index, to measure how strong should be the subtraction in the approach performed here. The fitting procedures indicate $\delta$-index value near to zero (a very weak dependence on the subtraction). This result allowed the use of a logarithmic representation for the subtraction. Then, the subtraction and the non-subtraction cases can be described by the same logarithmic Regge pole. 

The fitting procedures considering only energies above 1.0 TeV result in a pomeron intercept compatible with the double-pomeron value, $\alpha_{\mathbb{P}}\approx 1.04\sim1.05$. This value corroborates the Double-Logarithmic contributions, where the higher accuracy of calculations, the lower is the intercept, resulting in the best value is given by the intercept close to 1 \cite{B.I.Ermolaev.S.I.Troyan.Eur.Phys.J.C80.98.2020,B.I.Ermolaev.S.I.Troyan.Acta.Phys.Pol.B12.979.2019}.

The Regge cut in the original Regge formalism does not possess a clear role. However, in the present approach, the logarithmic Regge cut may represent the contributions coming from below the logarithmic Regge pole, $\alpha_{\mathbb{P}}$, when one adopts $\mathrm{disc}A(l,t)=(\alpha_c(t)-l)^{1+\beta(t)}$. Then, it can be used to explain the mixed region ($25.0 ~\mathrm{GeV} \leq \sqrt{s}\leq 1.0 ~ \mathrm{TeV}$), where the odderon and the pomeron compete as the leading contribution to the total cross section.

Then, one expects here that logarithmic Regge cut can describe the total cross section for $pp$ and $p\bar{p}$, from the minimum of the total cross section up to 1.0 TeV. However, assuming the parameters coming from the Regge cut can act as free fitting parameters, then they cause a supercritical value to the pomeron intercept, leading to the saturation of the FM bound. Notwithstanding, the dominance of the pomeron as the leading contribution at LHC energy seems to be an experimental fact \cite{L.Jenkovszky.R.Schicker.I.Szanyi.Int.J.Mod.Phys.E27.1830005.2018}. It is important to stress that the fitting procedures for the SET 1 and 2, using the parameterization (\ref{eq:sig_tot_1}), furnish an important constraint on the pomeron intercept $\alpha_{\mathbb{P}}$. Moreover, the experimental data at the cosmic-ray energies seem to have a little influence on the pomeron intercept. 

To solve this problem, it is necessary to use the experimental data for the $\rho$-parameter. In particular, one should use the experimental result at $\sqrt{s}=13.0$ TeV. This value seems to impose a double pomeron exchange, resulting in an intercept $\alpha_{\mathbb{P}}\leq 1$. When this experimental fact is used, then $\sigma_{tot}(s)$ rises below the saturation of FM bound. Of course, as stated in \cite{G.Antchev.etal.TOTEM.Coll.Eur.Phys.J.C79.785.2019}, the values for the $\rho$-parameter at $\sqrt{s}=13.0$ TeV excluded all the models classified and published by COMPETE. Therefore, the slowing down of the $\sigma_{tot}(s)$ seems to be given by the $\rho$-parameter at $\sqrt{s}=13.0$ TeV.

The slope of the differential cross section can also be predicted by the logarithmic Regge pole. Using the slope obtained by the TOTEM Collaboration, one predicts $B(\sqrt{s}=13.0~ \mathrm{TeV})=20.0\pm0.3$ GeV$^{-2}$, which is in accordance with the result obtained by the LHC, $B(\sqrt{s}=13.0~ \mathrm{TeV})=20.36\pm0.19$ GeV$^{-2}$.  

In the logarithmic Regge pole approach presented here, the Regge cut seems to has a clear role: it is responsible by the mixed region ($25.0~ \mathrm{GeV}\lesssim \sqrt{s}\lesssim 1.0~\mathrm{TeV}$) where the total cross section can be described by the odderon and the pomeron contributions. However, this result is strongly dependent on the discontinuities of the scattering amplitude. Unfortunately, there is no theoretical nor phenomenological information about discontinuity of $A(l,t)$.

\section*{Acknowledgments}

SDC thanks to UFSCar by the financial support. 


\end{document}